# Limit Order Book and its modelling in terms of Gibbs Grand-Canonical Ensemble

Alberto Bicci

Abstract

In the domain of the so called Econophysics some attempts already have been made for applying the theory of Thermodynamics and Statistical Mechanics to economics and financial markets. In this paper a similar approach is made from a different perspective, trying to model the limit order book and price formation process of a given stock by the Grand-Canonical Gibbs Ensemble for the bid and ask processes. As a consequence we can define in a meaningful way expressions for the temperatures of the ensembles of bid orders and of ask orders, which are a function of maximum bid, minimum ask and closure prices of the stock as well as of the exchanged volume of shares. It's demonstrated that the difference between the ask and bid orders temperatures can be related to the VAO (Volume Accumulation Oscillator) indicator, empirically defined in Technical Analysis of stock markets. Furthermore the distributions for aggregate bid and ask orders derived by the theory can be subject to well defined validations against real data, giving a falsifiable character to the model.

1. Introduction

In the last two decades physicists have tried to quantitatively model the behavior of financial markets with the tools offered by different branches of Physics, creating a new discipline called Econophysics (see for example [1]). Given the complex nature of the dynamics implied by financial markets, which is the result of the interactions between many actors and factors, many works have been concentrated on establishing analogies with the general laws of Thermodynamics (see for example [2]), putting in correspondence physical quantities such as energy, temperatures, pressure etc. with economical concepts. Of particular interest here is the possibility of applying the Gibbs theory of Canonical Ensembles, which are at the core of the statistical treatment of thermodynamics, to the behavior of financial markets ([3], [4]). As quoted in [4], "certainly the economy is a big statistical system with millions of participating agents and so presents itself as a possible and promising target for the application of statistical mechanics. All that remains is to find a sensible conserved quantity!".
Although economics is generally viewed as a non-equilibrium system, we believe, as stated in [3], that considering a relatively short period of time certain processes of financial markets can be modeled as an equilibrium system. This assumption is also at the center of our derivation. Assuming that a certain system is at equilibrium at least for a short period of time is a mandatory premise for the application of Gibbs Canonical Ensembles of Statistical Mechanics. Attempts of applying the different forms of Gibbs ensembles (micro-canonical, canonical and grand-canonical) to financial markets have been already tried in the past ([3]), trying to derive thermodinamical quantities like temperatures or entropy with a financial meaning. In [3] the ensembles are referred to the N economical agents of stock exchanges. This paper tries a different approach focusing on the process of price formation of a given stock. The starting point of our model is the order book and in particular the limit order book, which lists the aggregate proposals for buying and selling with a "limit" price a particular stock. As a matter of fact the order book configuration is at the origin of price determination. We don't try to consider the agents of the stock exchange as the constituents of our ensembles (i.e. the analogue of gas particles in Statistical Mechanics) but we treat the orders themselves (aggregate buy or sell orders) as the ensemble constituents. More precisely we model the process of price formation by interpreting the sets of aggregate bid (buy) and ask (sell) orders of a given stock as two separate gases to which we apply the Grand- Canonical Ensemble of Gibbs theory of thermodynamics. We take i.e. into consideration the limit order book of a stock and the statistics associated with the bid and ask orders.

A direct consequence of applying the Grand-Canonical Ensemble to the buy and sell orders is that they shall obey a well-defined mathematical law for the distribution of the average number of aggregate bid-ask orders as functions of proposed prices. Although the experimental validation has not been carried out in this work, the derived distribution law for bid-ask orders can be verified by studying the effective average distribution of order books in real stock exchanges. Said in other words the model is falsifiable. Another interest of the model is that it gives a not ambiguous and self-consistent meaning to the concept of temperature applied to financial markets and of other derived quantities such as the entropy or the energy. Furthermore a mathematical model for the order book distributions, which as stated above is at the origin of a stock price formation, is a potential tool for trying to predict the future prices. In fact it's shown in the following that an indicator empirically used in Technical Analysis for the prediction of future dynamics of stock prices, the VAO (Volume Accumulation Oscillator), is in direct relation with the difference of temperatures between the buy and sell orders ensembles.

The plan of the paper is as follows. In section 2 the concepts behind the possibility of using the theory of Gibbs Ensembles in our financial case are discussed. Section 3 applies the Grand-Canonical Ensembles to the bid and ask orders ensembles, deriving the concepts of temperatures. Section 4 exemplifies some practical calculation examples. Section 5 establishes the relation between the temperatures derived from the Grand-Canonical Ensembles with the VAO oscillator. Section 6 discusses some assumptions at the core of the derivation and highlights some possible future work.

## 2. Applicability of Statistical Mechanics to the Limit Order Book

The possibility of applying the concepts of Statistical Mechanics to financial market is not obvious; for a discussion see for example [11]. We must go back to the foundations to discern the applicability of such concepts. The main purpose of Statistical Mechanics is to compute the probability $W(n_\epsilon)$ of having $n_\epsilon$ particles of a given system in the states corresponding to energy levels $\varepsilon$. If we are able to establish a correct correspondence between the particles, the energy levels and the parameters which describe the financial market we can transpose many of the derivation of Statistical Mechanics to this domain. As better explained in the following we identify the energy levels with the possible prices that can be assumed by a stock in the phase of order book, i.e. in the phase in which buyers and sellers place proposals for buy (or bid in the financial jargon) and sell (or ask in the financial jargon). As the prices in stock negotiation are discretized (the quantum being the so-called thick), the "quantization" of energy levels introduced in Statistical Mechanics following the advent of Quantum Mechanics becomes natural in the transposition to financial markets. The role of the $n$ particles is identified in our model with the $n$ shares of a given stock which enter in the ask-bid negotiation of the order book.

A central concept in Statistical Mechanics is that of the phase space of spatial coordinates and momenta. A given energy level $\varepsilon$ corresponds generally to many states of the phase space. In [7] a phase space is defined for the coordinates of the market actors who buy and sell shares. In our specific case we didn't devise a reasonable way of introducing canonical coordinates and hence a phase space, but we think that not applying such concept doesn't destroy the possibility of applying the main concepts of the Gibbs theory to the financial markets.

In the Statistical Mechanics the concept of temperature arises from that of entropy: the temperature of a system at its equilibrium is given by $T = 1 / \frac{\partial S}{\partial E}$, where S is the entropy and E is the energy of the system.

The temperature is then a quantity derived from that of entropy, determining how the entropy varies when the total energy of the system varies. The entropy, at his turn, is defined as the logarithm of the number of states corresponding to a given energy E of the system, which is a concept generally applicable to complex systems and hence also to the financial markets. Since the entropy definition makes sense in the statistics of financial market, the temperature must also be a quantity full of meaning in this domain. It turns out as a matter of fact that the temperatures of bid and ask orders are well defined quantities which have a direct interpretation in terms of prices (energies) and number of exchanged shares (particles). The ensembles of bid orders and ask orders can be interpreted as two different gases at different temperatures. The difference between these temperatures is also a very interesting quantity, which we find to correspond to the empirical indicator VAO (Volume Accumulator Oscillator), currently used in the Technical Analysis of stock markets.

3. The Grand-Canonical Ensemble applied to the Limit Order Book

We take into consideration the limit order book of a given stock. The order book lists the proposals of public investors. There are two types of lists: one which lists the buy proposals (bid), detailing, for each price level, how many shares are proposed for purchase; the other which lists the sell proposals (ask), i.e., for each price how many shares are proposed for sale. These lists of course change continuously during time as the price of the stock evolves and the buy-sell proposals change accordingly. At each time interval, the last price of the stock is then determined as that price which maximizes the exchange (buy-sell) of shares. The table below reports an example of bid-ask order book.

| Price | Aggregate Bid Quantity | Aggregate Ask Quantity | Tradeable Quantity |
|---|---|---|---|
| $24.05 | 200 | 1,800 | 200 |
| $24.00 | 1,200 | 1,000 | 1,000 |
| $23.95 | 1,600 | 400 | 400 |

Tab. 1: Example of order book for a stock. The last price is $24,00 as it maximizes the tradeable quantity (1000 shares)

Generally the published order books list only a portion of the bid-ask proposals, for example the best 5 or 10 proposals. For the sake of our model we should consider the whole ensemble of proposals, although they are not easily publicly available.

On-book transactions represent only a portion, although the major one, of the trading. Part of the trading is done on the off-book market, where trades are arranged directly in a bilateral negotiation. It's however generally believed that on-book market is dominant in the price formation ([6]). Our model makes anyway reference solely to the mechanism of on-book price formation. We consider in addition limit order books only, i.e. orders which are entered with a "limit" price: buy (bid) limit orders are executed when the market price is below the threshold set by the investor and sell (ask) orders are executed when the market price is above the threshold set by the investor. The "aggregate" or "cumulative" bid (ask) quantity defined in Tab. 1 refers to the sum of limit orders with threshold below (above) the indicated price.

The starting point of our model is to compute the distributions of the bid and ask proposals. We want to determine the probability distribution $w(n_k^a)$ of having $n_k^a$ aggregate shares proposed for sale (ask) with the limit price $\epsilon_k$. In the same manner we want to compute the probability distribution $w(n_k^b)$ of having $n_k^b$ aggregate shares proposed for buy (bid) with the limit price $\epsilon_k$:

$w(n_k^a)$: probability of having an aggregate ask proposal of n shares above the "limit" price $\epsilon_k$

$w(n_k^b)$: probability of having an aggregate bid proposal of n shares below the "limit" price $\epsilon_k$

For the prices of the order proposals we consider the price difference between the proposed price $p(t)$ and the closure price of the stock at the previous time $\bar{p}(t - \Delta t)$, separated by a time interval $\Delta t$, (e.g. two consecutive days) $\varepsilon = p(t) - \bar{p}(t - \Delta t)$. We can similarly consider the log price difference as these don't differ from the latter for short time intervals. As we consider the price differences note that $\varepsilon$ may assume both positive and negative values. In the following we identify such price differences with the "energies" of the statistical ensembles with which we model the bid and ask order distributions. Differently from the Statistical Mechanics, where the energy levels are positive quantities, they can be both positive and negative in our financial model.

We can discretize the different values of prices between the minimum price proposed in the buy process $p_{min}$ (the minimum buy price) and the maximum price proposed in the sell process $p_{max}$ (the maximum ask price) into N levels, separated by a "quantum" difference that could be identified with the price tick used in trading (we implicitly assume that the maximum ask price is higher than the minimum bid price). As we consider for the "energy" levels at time t the price difference with the prices at time $(t - \Delta t)$ we must as well identify the minimum energy level at time t with $\varepsilon_{min} = p_{min}(t) - \bar{p}(t - \Delta t))$ and the maximum energy level $\varepsilon_{max} = p_{max}(t) - \bar{p}(t - \Delta t)$, where $\bar{p}(t - \Delta t)$ is again the last closure price.

There is no particular assumption on the time interval $\Delta t$ to be considered. The time interval should be however long enough for the distributions $w(n_k^a)$ and $w(n_k^b)$ of the order proposals to be statistically significant both for the bid and ask process. It should be short enough for considering the distribution as "stable", i.e. in an "equilibrium" state at the end of this time interval, when the price is defined.

In statistical mechanics the probability for a gas with a variable number N of particles of having $n_k$ particles in the energy level $\epsilon_k$ is given by the Grand-Canonical Ensemble of Gibbs ([8]). In our analogy the number of particles corresponds to the number of shares proposed for trading and the energy levels to price difference involved in a trading session and in particular in the bid-ask process. As the total number of shares proposed in the bid-ask process changes continuously the relevant ensemble is the Grand-Canonical one. Then for the bid (buy) proposals we have:

$$w(n_k^b) = \exp(\frac{\Omega_k^b + (\mu^b - \epsilon_k)n_k}{T^b}) \qquad (1)$$

where

$$\Omega_k^b = -T^b \ln \sum_k [\exp\left(\frac{\mu^b - \epsilon_k}{T^b}\right)]^{n_k} \qquad (2)$$

Here $T^b$ is the "temperature" of the bid ensemble (adopting the natural units in which the Boltzmann constant K=1) and $\mu^b$ is the "chemical potential" of the bid ensemble. The meaning of the "chemical potential" in our financial application will be clarified later.

We can similarly derive the probability distribution for the "gas" corresponding to the ask (sell) proposals, as:

$$w(n_k^a) = \exp(\frac{\Omega_k + (\mu^a + \epsilon_k)n_k}{T^a}) \qquad (3)$$

where

$$\Omega_k^a = -T^a \ln \sum_k [\exp\left(\frac{\mu^a + \epsilon_k}{T^a}\right)]^{n_k} \qquad (4)$$

$T^a$ and $\mu^a$ are the corresponding "temperature" and "chemical potential" for the "ask gas". As the "ask gas" is not in equilibrium with the "bid gas" the temperature and the chemical potential of the bid and ask gases will be in general different. Furthermore we assume that the "ask gas" energy levels $\epsilon_k$ are changed in sign (note the + sign in the exponent) with respect to the "bid gas". This makes sense as the bid orders prices are relevant to buy and the same prices in the ask orders are relevant to sell. This is a crucial assumption for the following derivation, without which the model shouldn't work, as the two distributions would have not intersection point. We stress the fact that the two ask and bid gases are not interacting in our model. They are in different equilibrium states, at different temperatures. In practice this corresponds to the assumption that at each time interval the orders proposed by buyers and sellers are independent, influenced only by the previous closure price $\bar{p}(t - \Delta t)$.

Similarly to the derivation done in the Bose case in Statistical Mechanics, we can derive the expression for the potentials of equations (2) and (4):

$$\Omega_k^b = T^b \ln(1 - e^{\frac{\mu^b - \varepsilon_k}{T^b}}) \qquad (5)$$

$$\Omega_k^a = T^a \ln(1 - e^{\frac{\mu^a + \varepsilon_k}{T^a}}) \qquad (6)$$

In a way analogous, in Statistical Mechanics, to the determination of the average number of particles at a given energy level $\epsilon_k$ we can determine the average number of aggregate shares proposed at the limit price $\epsilon_k$ in the bid and ask processes as:

$$\bar{n}_k^b = -\frac{\partial \Omega_k^b}{\partial \mu_b} = \frac{1}{e^{\frac{\varepsilon_k - \mu_b}{T_b}} - 1} \qquad (7)$$

$$\bar{n}_k^a = -\frac{\partial \Omega_k^a}{\partial \mu_a} = \frac{1}{e^{-\frac{\varepsilon_k + \mu_a}{T_a}} - 1} \qquad (8)$$

We stress the fact that these distributions give the average number of shares proposed for each price. Like the thermodynamic case where the real number of particles in a given energy level can fluctuate around the average value, the same shall apply to the bid and ask proposals in the financial markets. Expressions (7) and (8) should then be seen as an "approximation" or an average of the possible real distributions of the number of bid and ask shares as a function of price.

The figure 1 reports the typical behavior of the distribution for the two gases. As evidenced in the figure the assumption of assigning opposite sign to the energy levels of the bid ask with respect to the ask gas is mandatory for the two curves to always have an intersection point.

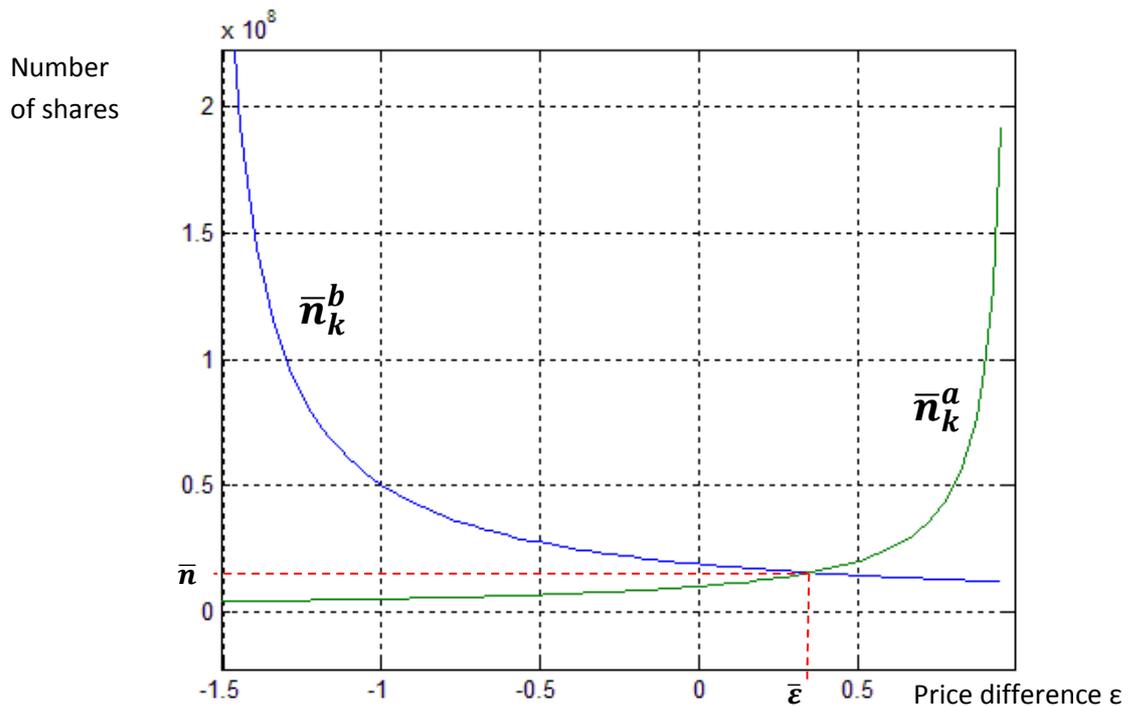

**Fig. 1: distribution law for the average number of aggregate shares proposed for buy (bid) and for sell (ask) as a function of the limit price difference $\varepsilon$**

As already stated a thorough comparison between the model and real distributions of bid and ask processes has not been carried out in this work, but some data reported in current literature seem to not contradict the "exponential" character of the distributions implied by equations (7) and (8). Figure 2, taken from [9], reporting the histograms of aggregate bid and ask orders of the stock CSCO on a given day, exhibit visually an exponential nature.

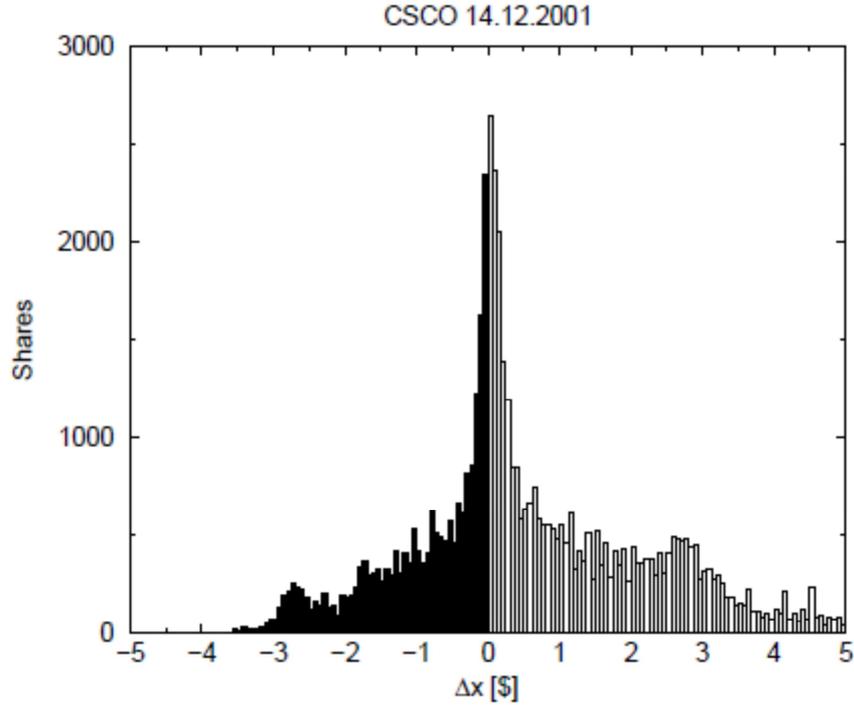

**Fig. 2: Histogram of aggregate orders of CSCO on 14.12.2000. The left distribution is the sell (ask) side and the right distribution is the buy (bid) side. Here $\Delta x$ represents respectively the difference between the proposed price and the maximum sell price (left) and the difference between the proposed price and the minimum bid (right).**

Also [10] reports some distributions for limit order books which, although with some deviations, confirm a substantially exponential decay character.

As the bid curve tends asymptotically to the value of the chemical potential $\mu_b$ (in statistical mechanics the chemical potential is always negative) it's natural to identify $\mu_b$ with the minimum ask price $\varepsilon_{min} = p_{min}(t) - \bar{p}(t - \Delta t)$. Similarly the ask curve tends asymptotically to the value $-\mu_a$, which can be identified with the maximum energy-price level $\varepsilon_{max} = p_{max}(t) - \bar{p}(t - \Delta t)$. Assuming that the values of $\mu_a$ and $\mu_b$ are negative corresponds to the assumption that the minimum price offered in the bid process $p_{min}(t)$ is always lower than the previous closure price $\bar{p}(t - \Delta t)$ and that the maximum price offered in the ask process is always higher than the previous closure price $\bar{p}(t - \Delta t)$. The identification of chemical potential with price, although in a different context, has already been done in [2].

In the bid-ask process the resulting stock price at a given time is the one that maximizes the exchange between demand and offer, i.e. the price at which the number of shares offered for sale is closest to that asked for purchase. This condition, in our model, is approximated by the condition

$$\bar{n}_k^b = \bar{n}_k^a = \bar{n} \qquad (9)$$

The price $\bar{\epsilon}$ which fulfills this condition corresponds to the intersection point of bid and ask exponential curves in fig. 2.

Equation (9) results in the determination of the new closure price at time t and of the exchanged share volume:

$$\bar{\epsilon} = \frac{\mu_a T_a - \mu_b T_b}{T^a - T^b} \quad (10)$$

$$\bar{n} = \frac{1}{e^{\frac{\mu_a + \mu_b}{T^a + T^b}} - 1} \quad (11)$$

If then we know the minimum bid price and the maximum ask price we can derive the temperatures $T^b$ and $T^a$ of bid and ask orders from the knowledge of the resulting asset price $\bar{\epsilon}$ (e.g the closure price at time t) and of the exchanged volume of the asset $\bar{n}$, resulting in:

$$T^b = \frac{\mu_b + \bar{\epsilon}}{-\ln(1 + \frac{1}{\bar{n}})} \quad (12)$$

$$T^a = \frac{\mu_a - \bar{\epsilon}}{-\ln(1 + \frac{1}{\bar{n}})} \quad (13)$$

These are the temperatures of the bid and ask "gases", i.e. the temperatures of the buyers and sellers order ensembles which give rise to the price $\bar{\epsilon}$ and to the exchange of $\bar{n}$ shares.

Putting $\bar{\epsilon} = \bar{p}(t) - \bar{p}(t - \Delta t)$ and $-\mu_a = p_{max}(t) - \bar{p}(t - \Delta t)$ ; $\mu_b = p_{min}(t) - \bar{p}(t - \Delta t)$, the expressions for $T^b$ and $T^a$ can be recast as ($\bar{p}(t) = \bar{\epsilon}$):

$$T^b = \frac{p_{max}(t) - \bar{p}(t)}{\ln(1 + \frac{1}{\bar{n}})} \quad (14)$$

$$T^a = \frac{\bar{p}(t) - p_{min}(t)}{\ln(1 + \frac{1}{\bar{n}})} \quad (15)$$

This makes evident that $T^b$ and $T^a$ are both positive quantities, as it should be.

We can also compute the difference $\Delta T = T^a - T^b$ between the sell and buy gases, resulting in :

$$\Delta T = T^a - T^b = \frac{2\left[\frac{p_{min}(t) + p_{max}(t)}{2} - \bar{p}(t)\right]}{\ln(1 + \frac{1}{\bar{n}})} \quad (16)$$

As we'll see in the following this temperature difference is strictly related to the VAO oscillator used in the Technical Analysis of financial markets.

Other thermodinamical quantities can of course be derived that can be of interest for our financial case. The internal energy of ask and bid gases can be determined, as well as the entropy S.

4. **Practical computation**

The computation of $T^b$ and $T^a$ implies the knowledge of the closure price $\bar{p}(t)$ and $\bar{p}(t - \Delta t)$, of the exchanged volume of asset $\bar{n}$ and of the maximum ask price $p_{max}(t)$ and the minimum bid price $p_{min}(t)$ registered in the limit order book in the time period $\Delta t$ considered in our process. While both the closure price and the exchanged volume of a stock can be easily collected from public data, the retrieval of $p_{max}(t)$ and $p_{min}(t)$ is definitely more difficult.

For easing the computation of $T^b$ and $T^a$ based on available public data we can approximate the value of $p_{max}(t)$ and $p_{min}(t)$ with the maximum and minimum price of the stock reached during the period of time $\Delta t$. This is clearly not fully correct and provides an underestimation of the maximum ask price $p_{max}(t)$ as well as an overestimation of the minimum bid price $p_{min}(t)$.

Operating in this way we can easily compute $T^b$ and $T^a$, based on the equations (14) and (15), where we substitute $p_{max}(t)$ with $\tilde{p}_{max}(t)$ and $p_{min}(t)$ with $\tilde{p}_{min}(t)$.

Fig. 3 reports the historical data between Nov. 1st 2014 and Nov. 1st 2015 for the ENI stock (FTSE MIB stock exchange), together with the temperatures $T^b$ and $T^a$ and the difference $\Delta T = T^a - T^b$, where we have chosen for $\Delta t$ 1 day.

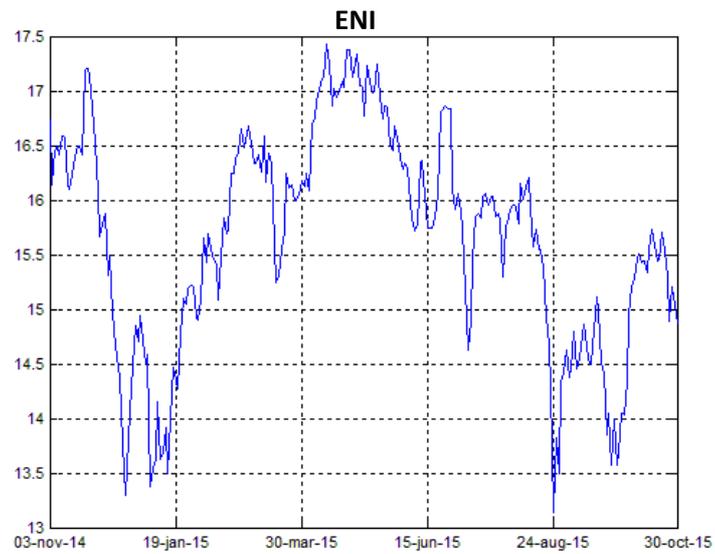

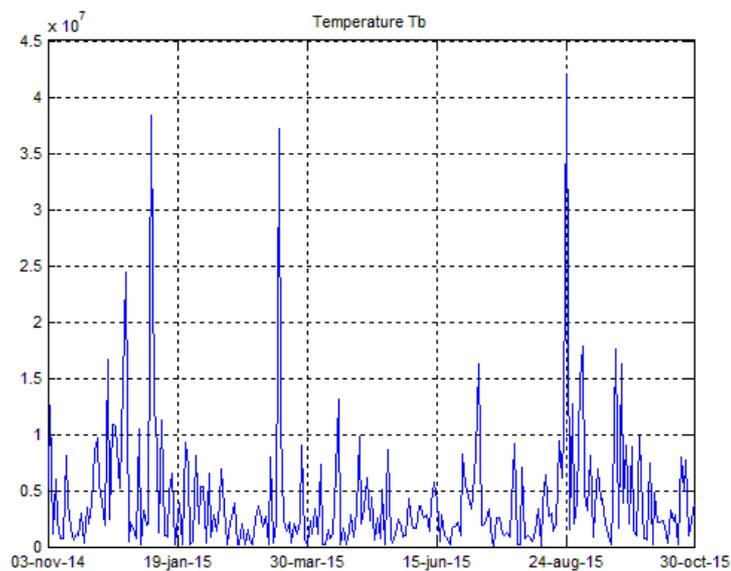

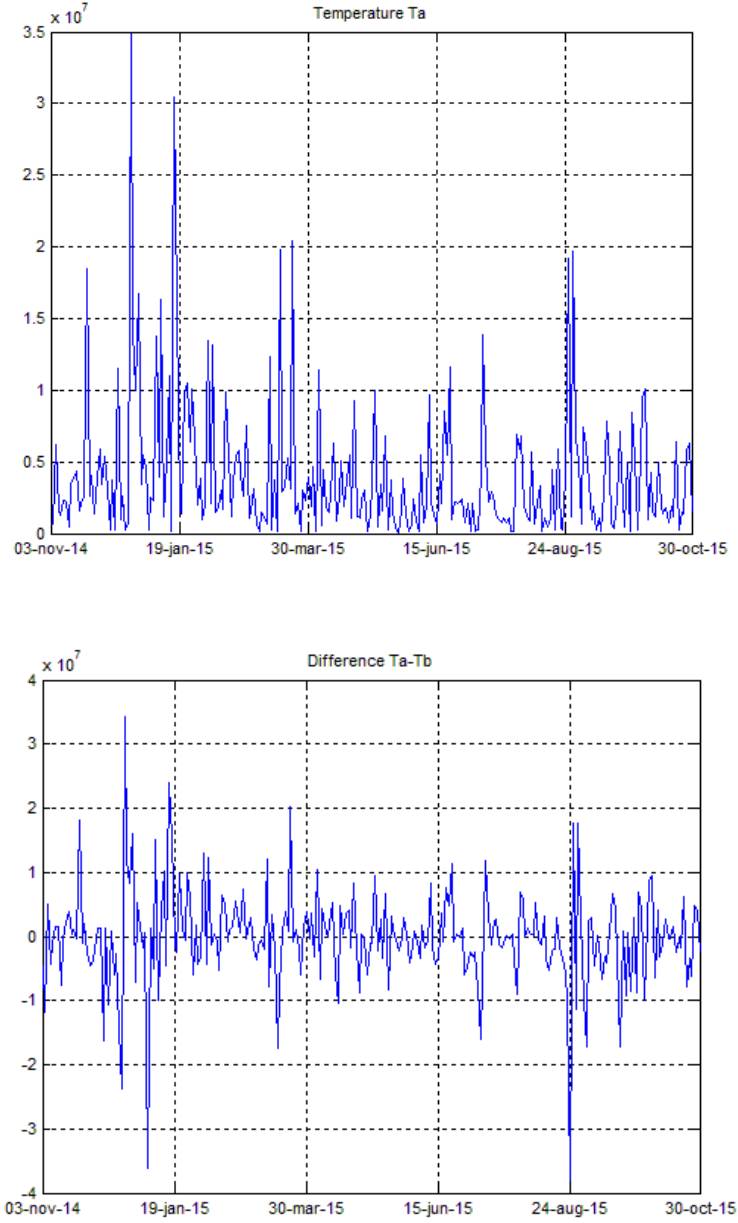

Fig. 3: Price data for the ENI stock between Nov. 1st 2014 and Nov 1st 2015 and the computed temperatures $T^b$, $T^a$ and $\Delta T$.

## 5. Correspondence with the VAO oscillator

In equations (14), (15) and (16) the volume of exchanged shares $\bar{n} \gg 1$ so that we can approximate the natural logarithm in the denominator with its first order expansion, resulting in:

$$T^b = \bar{n}[\tilde{p}_{max}(t) - \bar{p}(t)] \qquad (17)$$

$$T^a = \bar{n}[\bar{p}(t) - \tilde{p}_{min}(t)] \qquad (18)$$

$$\Delta T = 2\bar{n}\left[\frac{\tilde{p}_{min}(t) + \tilde{p}_{max}(t)}{2} - \bar{p}(t)\right] \qquad (19)$$

where, as discussed in Section 4, we have used the maximum and minimum stock prices instead of the maximum ask and minimum bid.

In the technical analysis of financial market one often used oscillator is the VAO (Volume Accumulation Oscillator) [5]. The VAO oscillator, computed over a moving time window (of for example N days) is defined as:

$$VAO = \sum_{i=1}^{N} 2\bar{n} \left[ \frac{\tilde{p}_i^{min} + \tilde{p}_i^{max}}{2} - \bar{p}_i \right]$$

We see then that the accumulation of the quantity $\Delta T$ in (19) over a time window of N days is exactly the VAO.

The VAO, which is a quantity derived in a purely empirical manner within the Technical Analysis, finds therefore an explanation in our thermodinamical model as the difference of "temperatures" between the sell and buy orders. This opens the possibility of exploiting the thermodynamics quantities for the bid and ask gases derived from our model (not only the temperatures but also the energies and/or the entropies) as further predictors of future financial market trends.

6. Discussion

Our theory basically models the distribution of the aggregate buy and sell orders of a limit order book, assuming that, given a time period $\Delta t$, we can state that these distributions are at equilibrium and follow the laws of Statistical Mechanics for a Bose gas. We can therefore derive the <u>average</u> distributions for the aggregate buy and sell orders as those governing the partition of the number of gas particles (the number of shares proposed for buy or sell in our case) in the different energy levels (the limit prices of buy or sell orders, considered as the price difference with respect to the previous closure price).

In such a way the resulting price and exchanged volume are then obtained as the price and volume coordinates of the intersection between the two distributions, as this is the price at which the maximum volume of shares is exchanged between buyers and sellers.

Once we have made the assumption of considering the buy and sell orders (separately) as gases, the use of the powerful Gibbs ensemble from Statistical Mechanics does nothing but deriving a mathematical expression for the average distribution of those orders.

A further study, which has not been carried out in this work, could make a statistical analysis of the limit order books for real stocks, for verifying if the average distributions for aggregate buy and sell orders are effectively well approximated by the Gibbs distributions, reported in equations (7) and (8). This could be a good verification of the hypotheses and derivations made in this paper.

As previously stated one crucial assumption at the base of the derivation presented in this paper is the equilibrium hypothesis for the buy and sell gases. The Grand-Canonical Ensemble applies only to gases in equilibrium condition. As it's well known the limit order book is continuously evolving in time: new proposal of buy and sell are continuously entering (without considering the order cancellations) and the determination of price which maximizes the exchange, i.e. the price of the stock under consideration, evolves consequently. Our assumption is that, provided that we can choose an appropriate time interval $\Delta t$, the time evolution of limit order book can be seen as a succession of equilibrium states, so that for each of them we can define a temperature for the ensembles of buy and sell orders.

For the appropriate time interval $\Delta t$ we should consider the ensembles of buy and sell orders which have been proposed in that time interval as the ones to which apply the Gibbs statistics. A further study aimed at analyzing the effective distributions of buy and sell orders and comparing them with the model presented in this paper should also investigate the time interval for which this hypothesis is valid.

We must also remark the fact that our distribution model is surely not exact, as for example is not realistic the fact that the distribution of equation (7) and (8) assume infinite values for $\varepsilon = \mu_b$ (the minimum bid price) and $\varepsilon = -\mu_a$ (the maximum ask price). Further studies, corroborated also from an analysis of the behavior of real distributions, should probably try to infer a more correct distribution, without these infinities.

## 7. Conclusion

The paper presents a theory of the distributions for buy and sell limit order books based on an analogy with the Grand-Canonical Ensemble of Gibbs theory of thermodynamics. The theory appears self-consistent and permits the derivation of temperatures (always positive) of the sell and buy orders ensembles. Treating them as "gases" permits also the definition of other thermodinamical quantities such as the internal energy and the entropy.

The accumulated difference between the temperatures of ask and bid order distributions appears to be exactly the VAO oscillator empirically defined in Technical Analysis, if maximum and minimum stock prices are used instead of the maximum ask and minimum bid, which enter in the original equations for the temperatures.

The model provides a law for the average distributions of aggregate bid and ask orders which is subject to an experimental validation, through an analysis of the real distributions in limit order books. This comparison should also investigate whether the time interval over which to calculate the distributions affects the result and what is the optimal interval for the theoretical distribution to effectively approximate the real one.

On the other hand some inconsistency in the model, such as the asymptotic values of the distributions, lead to believe that the distributions derived from the Grand-Canonical Ensemble cannot be fully representative of the real ones. However we believe that this simple model could foster further research for achieving a more exact distribution model, based also on the aforementioned verification activity on real data.